\documentstyle[onecolumn,epsf]{mn}

\newcommand{\sD}{{\cal D}}
\newcommand{\sL}{{\cal L}}

\newcommand{\bd}{\mbox{\boldmath$d$}}

\newcommand{\bff}{\mbox{\boldmath$f$}}
\newcommand{\br}{\mbox{\boldmath$r$}}

\title[Vertical resonances in accretion discs]
{A non-linear theory of vertical resonances in accretion discs}

\author[G. I. Ogilvie]
  {G. I. Ogilvie\\
  Institute of Astronomy, University of Cambridge, Madingley Road,
  Cambridge CB3 0HA}

\begin{document}

\maketitle

\label{firstpage}

\begin{abstract}
  An important and widely neglected aspect of the interaction between
  an accretion disc and a massive companion with a coplanar orbit is
  the vertical component of the tidal force.  As shown by Lubow, the
  response of the disc to vertical forcing is resonant at certain
  radii, at which a localized torque is exerted, and from which a
  compressive wave (p mode) may be emitted.  Although these vertical
  resonances are weaker than the corresponding Lindblad resonances,
  the $m=2$ inner vertical resonance in a binary star is typically
  located within the tidal truncation radius of a circumstellar disc.
  
  In this paper I develop a general theory of vertical resonances,
  allowing for non-linearity of the response, and dissipation by
  radiative damping and turbulent viscosity.  The problem is reduced
  to a universal, non-linear ordinary differential equation with two
  real parameters.  Solutions of the complex non-linear Airy equation
  are presented to illustrate the non-linear saturation of the
  resonance and the effects of dissipation.  It is argued that the
  $m=2$ inner vertical resonance is unlikely to truncate the disc in
  cataclysmic variable stars, but contributes to angular momentum
  transport and produces a potentially observable non-axisymmetric
  structure.
\end{abstract}

\begin{keywords}
  accretion, accretion discs -- binaries: close -- celestial mechanics
  -- hydrodynamics -- waves.
\end{keywords}

\section{Introduction}

\subsection{Horizontal and vertical tidal forces}

Accretion discs in binary stars and protoplanetary systems are subject
to periodic tidal forcing from the orbiting companion objects.
Similar effects can occur in planetary ring systems, where the
planetary satellites provide the perturbing forces, and in galactic
discs.  Even in the simple case of a companion with a circular orbit
that is coplanar with the disc, the effects of the tidal force are
manifold.

Consider a system of two spherical bodies of masses $M_1$ and $M_2$ in
a bound orbit.  Consider a third, test body orbiting about $M_1$.  In
a non-rotating coordinate system centred on $M_1$, the force per unit
mass experienced by the test body is
\begin{equation}
  \bff=-{{GM_1\,\br}\over{|\br|^3}}-
  {{GM_2(\br-\bd)}\over{|\br-\bd|^3}}-
  {{GM_2\,\bd}\over{|\bd|^3}},
\end{equation}
where $\bd(t)$ is the position vector of the companion $M_2$ with
respect to $M_1$, and the third term is the fictitious force arising
from the acceleration of the origin of the coordinate system.
Inasmuch as the second and third terms perturb the Keplerian motion of
the test body about $M_1$, they represent the tidal force in this
system.

Consider a thin accretion disc around $M_1$, and let $(R,\phi,z)$ be
cylindrical polar coordinates such that $z=0$ is the mid-plane of the
disc.  Suppose that the companion has a circular, coplanar orbit of
radius $d$ and angular velocity $\omega$.  Then the components of the
tidal acceleration are
\begin{equation}
  f_{{\rm t}R}=-GM_2[R^2+d^2-2dR\cos(\phi-\omega t)+z^2]^{-3/2}
  [R-d\cos(\phi-\omega t)]-GM_2d^{-2}\cos(\phi-\omega t),
\end{equation}
\begin{equation}
  f_{{\rm t}\phi}=-GM_2[R^2+d^2-2dR\cos(\phi-\omega t)+z^2]^{-3/2}
  d\sin(\phi-\omega t)+GM_2d^{-2}\sin(\phi-\omega t),
\end{equation}
\begin{equation}
  f_{{\rm t}z}=-GM_2[R^2+d^2-2dR\cos(\phi-\omega t)+z^2]^{-3/2}z.
\end{equation}

The horizontal components of the tidal acceleration are nearly
independent of $z$ within the disc and cause a general
non-axisymmetric distortion of its streamlines and surface density
distribution (Papaloizou \& Pringle 1977).  Indeed, the streamlines of
the disc correspond closely to a family of simple periodic orbits of
the restricted three-body problem found by Paczy\'nski (1977).  The
tidal distortion is enhanced in the neighbourhood of Lindblad
resonances, radii at which the forcing frequency resonates with the
natural epicyclic oscillations of the disc.  Usually at a Lindblad
resonance, a non-axisymmetric wave will be launched and will then
propagate some way radially through the disc before dissipating and
transferring its angular momentum to the disc (Goldreich \& Tremaine
1979).  The resonant torque exerted between the companion and the disc
may be calculated from a standard formula that applies under quite
general circumstances in linear theory, for example if the response is
dominated by viscosity and no wave is emitted (Meyer-Vernet \& Sicardy
1987), or if the vertical structure of the disc is taken fully into
account (Lubow \& Ogilvie 1998).

Tidal torques usually limit the outer radius of a circumstellar disc
(or the inner radius of a circumbinary disc).  Except in the case of a
very low-mass companion, the Lindblad resonances are so strong that
they are excluded from the disc.  In a binary star with a mass ratio
$q=M_2/M_1$ of order unity, a circumstellar disc is tidally truncated
well inside the innermost ($m=2$) Lindblad resonance, which in any
case lies outside the Roche lobe (Paczy\'nski 1977; Papaloizou \&
Pringle 1977).  For smaller mass ratios typical of giant planets
orbiting stars, the Lindblad resonances play a more direct role in
truncating the disc (Lin \& Papaloizou 1986).

Another horizontal tidal effect is the eccentric interaction between
the companion and the disc.  At eccentric Lindblad resonances such as
the 3:1 resonance, a local eccentric instability occurs, which may be
able to sustain a large-scale eccentric distortion of the disc (Lubow
1991).  The eccentric corotational resonances act to damp
eccentricity, however.

Studies of the {\it vertical\/} component of the tidal force have
mostly been concerned with the case of a companion with an inclined
orbit.  In that case, the vertical tidal acceleration is nearly
independent of $z$ within the disc.  As a result, non-axisymmetric
bending waves are launched at vertical resonances where the forcing
frequency resonates with the natural vertical oscillations of the disc
(Shu, Cuzzi \& Lissauer 1983).  In a Keplerian disc the Lindblad and
vertical resonances coincide.

For a companion with a {\it coplanar\/} orbit, however, the vertical
tidal acceleration is proportional to $z$ and vanishes on the
mid-plane.  Other than those concerned with bending waves, most
analytical and numerical studies have relied on a two-dimensional
treatment of the disc, and the vertical component of the tidal force
has been widely neglected.  However, Lubow (1981) analysed the
response of a vertically isothermal disc to vertical tidal forcing by
a companion with a circular, coplanar orbit.  The general response
consists of a non-axisymmetric vertical expansion and contraction of
the disc, characterized by a vertical velocity proportional to $z$.
This `breathing' motion corresponds to a compressive mode of the disc,
the $n=1$ p mode in the notation of Lubow \& Pringle (1993), or the
${\rm p}_1^{\rm e}$ mode in the classification of Ogilvie (1998).  At
particular radii where the forcing frequency resonates with the
natural frequency of this mode, a non-axisymmetric p-mode wave may be
launched and a resonant torque exerted.  These are also {\it vertical
  resonances}, but are distinct from those associated with bending
waves because a different mode is involved.  Lubow's vertical
resonances are further from the companion than the corresponding
Lindblad resonances and, in a thin disc, the resonant torques are much
weaker.  However, Lubow (1981) pointed out that, in a binary star with
a mass ratio of order unity, the $m=2$ inner vertical resonance
typically lies within the standard tidal truncation radius, and the
associated torque can compete with the viscous torque in the disc.

Another consequence of the vertical tidal force is a local tilt
instability, closely related to the eccentric instability (Lubow
1992).  The tilt instability occurs at inclination resonances that are
essentially coincident with the eccentric Lindblad resonances, but the
tilt instability is much weaker.

\subsection{Aims of the paper}

In this paper I develop a non-linear theory of Lubow's vertical
resonances.  I generalize the analysis of Lubow (1981) to
non-isothermal discs allowing for non-linearity, viscosity and
radiative damping.  By reducing the problem to a universal
differential equation and studying its solutions, I aim to demonstrate
the effects of non-linear saturation and dissipation on the resonant
response of differentially rotating discs to periodic forcing in
general.

In a related paper (Ogilvie 2001, hereafter Paper~I) I have presented
an analysis of the non-linear tidal distortion of a thin,
three-dimensional accretion disc by a binary companion on a circular
orbit.  I showed that the resulting distortion can plausibly explain
the two-armed non-axisymmetric features seen in the Doppler tomograms
of IP~Peg and other dwarf novae in outburst (e.g. Steeghs 2001).
Although the $m=2$ inner vertical resonance plays an important role in
regulating the amplitude and phase of the tidal distortion, the method
of analysis in Paper~I does not allow for the emission of a wave from
the vertical resonance.  One of the aims of the present paper is to
justify that limitation by showing that, under typical conditions,
vertical resonances are broadened and damped by non-linearity and
dissipation to the extent that the emission of a wave may reasonably
be neglected.

Before embarking on the detailed analysis, it may be helpful to look
ahead to the equation to be derived,
\begin{equation}
  -{{{\rm d}^2y}\over{{\rm d}x^2}}+xy+|y|^2y+{\rm i}by=a.
\end{equation}
In this equation, $x$ represents the radial distance from the resonant
orbit and $y$ the complex amplitude of the response, both in
dimensionless terms.  The linear terms $-y''+xy$ represent the ability
of the disc to support freely propagating waves in the region $x<0$.
In the present case these waves correspond to the compressive ${\rm
  p}_1^{\rm e}$ mode.  The point $x=0$ is the location of the
resonance, which is also the turning point of free waves.  The
constant term $a$ on the right-hand side represents the tidal forcing
in the neighbourhood of the resonance.  The inhomogeneous Airy
equation $-y''+xy=a$ is familiar in studies of resonant wave
excitation in differentially rotating discs and features in the
analysis of Lubow (1981).  The new, non-linear term $|y|^2y$ derives
from couplings between the resonant mode and non-resonant modes.  It
leads to new effects such as the non-linear broadening and saturation
of the resonance.  The correct derivation of this term accounts for
much of the complexity of the analysis in the present paper.  The
linear term ${\rm i}by$ represents the dissipation of vertical motions
by shear and bulk viscosity and by radiative damping.  It leads to
attenuation of the waves and broadening of the resonance.

The remainder of this paper is organized as follows.  In Section~2 I
recall the coordinate system and the basic equations used in Paper~I.
In Section~3 I expand the equations in the neighbourhood of the
vertical resonance and solve them in a systematic manner to derive the
complex non-linear Airy equation.  In Section~4 I discuss the
properties of the equation and present numerical solutions for a range
of parameter values.  The astrophysical consequences of the analysis
are explored in Section~5.

\section{Analysis}

\label{analysis}

\subsection{Orbital coordinates}

Consider a binary system with two stars in a circular orbit.  The
stars are approximated as spherical masses $M_1$ and $M_2$.  Let $d$
be the binary separation and
\begin{equation}
  \omega=\left[{{G(M_1+M_2)}\over{d^3}}\right]^{1/2}
\end{equation}
the binary orbital frequency.  Consider the binary frame rotating
about the centre of mass with angular velocity $\omega$, and let
$(R,\phi,z)$ be cylindrical polar coordinates such that star 1 (about
which the disc orbits) is at the origin, while star 2 has fixed
coordinates $(d,0,0)$.

In Paper~I a system of non-orthogonal {\it orbital coordinates\/}
($\lambda,\phi,z$) was introduced instead of cylindrical polar
coordinates.  These are based on Paczy\'nski's orbits for the
restricted three-body problem, and naturally account for the principal
tidal distortion of the disc.  The coordinate $\lambda$ is a
quasi-radial coordinate that labels the orbits and differs only
slightly from $R$, reducing exactly to $R$ in the limit $M_2\to0$.  In
this paper I again consider the case of a prograde circumstellar disc.
However, the generalization to retrograde and circumbinary discs is
straightforward.

The standard notation of tensor calculus is adopted, with $g_{ab}$
being the metric tensor, $J=\sqrt{\det(g_{ab})}$ the Jacobian of the
coordinate system and $\eta^{abc}$ the Levi-Civita alternating tensor.
Expressions for the metric coefficients and connection components for
the orbital coordinate system may be found in Paper~I, Section~2.

\subsection{Basic equations}

\label{basic}

The basic equations governing a fluid disc in three dimensions are the
same as in Paper~I.  However, the effect of a non-zero relaxation time
of the turbulent stress will not be considered here, as it was found
to introduce algebraic complications without affecting the final result
of this paper significantly.  Accordingly, a purely viscous model of
the turbulent stress is adopted here.

The equation of mass conservation is
\begin{equation}
  (\partial_t+u^a\nabla_a)\rho=-\rho\nabla_au^a,
  \label{drho}
\end{equation}
where $\rho$ is the density and $u^a$ the velocity relative to the
rotating frame.  The equation of motion is
\begin{equation}
  \rho\left[(\partial_t+u^b\nabla_b)u^a+2\eta^{abc}\omega_bu_c\right]=
  -\rho\nabla^a(\Phi+\Phi_{\rm t})-\nabla^ap+\nabla_bT^{ab}.
  \label{du}
\end{equation}
Here $\omega^a$ is the angular velocity of the rotating frame (only
$\omega^z=\omega$ being non-zero), $p$ is the pressure,
\begin{equation}
  \Phi(R,z)=-GM_1(R^2+z^2)^{-1/2}-{{1}\over{2}}\omega^2R^2
\end{equation}
is the part of the effective potential not due to star 2, and
\begin{equation}
  \Phi_{\rm t}(R,\phi,z)=-GM_2(R^2+d^2-2dR\cos\phi+z^2)^{-1/2}+
  GM_2d^{-2}R\cos\phi
\end{equation}
is the tidal potential due to star 2.  The turbulent stress tensor
$T^{ab}$ is assumed to be given by
\begin{equation}
  T^{ab}=\mu(\nabla^au^b+\nabla^bu^a)+
  (\mu_{\rm b}-{\textstyle{{2}\over{3}}}\mu)(\nabla_cu^c)g^{ab},
  \label{tab}
\end{equation}
where $\mu$ is the effective shear viscosity and $\mu_{\rm b}$ the
effective bulk viscosity.  The energy equation is
\begin{equation}
  \left({{1}\over{\gamma-1}}\right)(\partial_t+u^a\nabla_a)p=
  -\left({{\gamma}\over{\gamma-1}}\right)p\nabla_au^a+
  {T^{ab}}\nabla_au_b-\nabla_aF^a,
  \label{dp}
\end{equation}
where $\gamma$ is the adiabatic exponent and $F^a$ the radiative
energy flux, given in the Rosseland approximation for an optically
thick medium by
\begin{equation}
  F^a=-{{16\sigma T^3}\over{3\kappa\rho}}\nabla^aT,
\end{equation}
where $\sigma$ is the Stefan-Boltzmann constant, $T$ the temperature
and $\kappa$ the opacity.  The equation of state of an ideal gas,
\begin{equation}
  p={{k\rho T}\over{\mu_{\rm m}m_{\rm H}}},
\end{equation}
is adopted, where $k$ is Boltzmann's constant, $\mu_{\rm m}$ the mean
molecular weight and $m_{\rm H}$ the mass of the hydrogen atom.  The
opacity is assumed to be of the power-law form
\begin{equation}
  \kappa=C_\kappa\rho^xT^y,
\end{equation}
where $C_\kappa$ is a constant.  This includes the important cases of
Thomson scattering opacity ($x=y=0$) and Kramers opacity ($x=1$,
$y=-7/2$).

The effective viscosity coefficients are assumed to be given by an
alpha parametrization.  The precise form of the alpha prescription
relevant to a tidally distorted disc is to some extent debatable.  It
will be convenient to adopt the form
\begin{equation}
  \mu=\alpha p\left({{GM_1}\over{\lambda^3}}\right)^{-1/2},\qquad
  \mu_{\rm b}=\alpha_{\rm b}p\left({{GM_1}\over{\lambda^3}}\right)^{-1/2},
\end{equation}
where $\alpha$ and $\alpha_{\rm b}$ are the dimensionless shear and
bulk viscosity parameters.  In the limit of a circular disc in the
absence of star 2, this prescription reduces to the usual one,
$\mu=\alpha p/\tilde\Omega$, etc., where $\tilde\Omega$ is the orbital
angular velocity in the inertial frame.  For simplicity, the five
dimensionless parameters of the disc, $(\alpha,\alpha_{\rm
  b},\gamma,x,y)$ will be assumed to be constant throughout this
paper.  The other parameter in the problem is the mass ratio
$q=M_2/M_1$.

\section{Expansions}

\subsection{Thin disc}

Analytical progress can be made only if the disc is thin so that
certain approximations may be used.  The systematic way of introducing
these approximations is through an asymptotic analysis.

Let a system of units be adopted such that the radius of the disc and
the orbital frequency are $O(1)$.  Let the small parameter $\epsilon$
be a characteristic value of the angular semi-thickness $H/R$ of the
disc.  To resolve the vertical structure of the disc, introduce a
stretched vertical coordinate $\zeta$ according to
\begin{equation}
  z=\epsilon\zeta,
\end{equation}
so that $\zeta=O(1)$ within the disc.

Any slow time-dependence of the unperturbed disc, which occurs on a
characteristic viscous time-scale
$\alpha^{-1}\epsilon^{-2}\Omega^{-1}$, is consistently neglected.  The
desired solution is stationary in the binary frame.

\subsection{Orbital motion}

Let
\begin{equation}
  \bar\Phi(R,\phi)=\Phi(R,0)+\Phi_{\rm t}(R,\phi,0)
\end{equation}
denote the total potential {\it in the binary plane}.  The equations
governing the shape $R(\lambda,\phi)$ of Paczy\'nski's orbits and
their orbital angular velocity $\Omega(\lambda,\phi)$ can be expressed
in the form (Paper~I, Section~5.1)
\begin{equation}
  (R^2+2R_\phi^2-RR_{\phi\phi})\Omega^2+2(R^2+R_\phi^2)\omega\Omega=
  R\left({{\partial\bar\Phi}\over{\partial R}}\right)_\phi-
  {{R_\phi}\over{R}}\left({{\partial\bar\Phi}\over{\partial\phi}}\right)_R,
  \label{orbital_r}
\end{equation}
\begin{equation}
  \Omega\partial_\phi\left[R^2(\Omega+\omega)\right]=
  -\left({{\partial\bar\Phi}\over{\partial\phi}}\right)_R,
  \label{orbital_phi}
\end{equation}
where $R_\phi=(\partial R/\partial\phi)_\lambda$, etc.  These also
correspond to the horizontal components of the equation of motion of
the disc at leading order in $\epsilon$.  The divergence of the
orbital motion is
\begin{equation}
  \Delta={{1}\over{J}}\partial_\phi(J\Omega).
\end{equation}

The primary potential can be expanded in a Taylor series about the mid-plane,
\begin{equation}
  \Phi=\Phi^{(0)}(R)+{\textstyle{{1}\over{2}}}\epsilon^2\zeta^2
  \Phi^{(2)}(R)+O(\epsilon^4).
\end{equation}
The tidal potential can be expanded similarly, but is assumed formally
to be smaller by a factor $O(\epsilon)$ than the primary potential,
i.e.
\begin{equation}
  \Phi_{\rm t}=\epsilon\left[\Phi_{\rm t}^{(1)}(R,\phi)+
  {\textstyle{{1}\over{2}}}\epsilon^2\zeta^2\Phi_{\rm t}^{(3)}(R,\phi)+
  O(\epsilon^4)\right].
\end{equation}
Thus $\bar\Phi=\Phi^{(0)}+\epsilon\Phi_{\rm t}^{(1)}$, and Paczy\'nski's
orbits can be considered as deformed circular orbits such that the
orbital quantities possess expansions of the form
\begin{equation}
  R=\lambda+\epsilon R^{(1)}(\lambda,\phi)+O(\epsilon^2),
\end{equation}
\begin{equation}
  \Omega=\Omega^{(0)}(\lambda)+\epsilon\Omega^{(1)}(\lambda,\phi)+
  O(\epsilon^2),
\end{equation}
\begin{equation}
  \Delta=\epsilon\Delta^{(1)}(\lambda,\phi)+O(\epsilon^2),
\end{equation}
where
\begin{equation}
  \Omega^{(0)}=\left({{GM_1}\over{\lambda^3}}\right)^{1/2}-\omega.
\end{equation}
Let
\begin{equation}
  \tilde\Omega=\Omega^{(0)}+\omega=\left({{GM_1}\over{\lambda^3}}\right)^{1/2}
\end{equation}
denote the angular velocity in the inertial frame.

An explicit calculation of $R^{(1)}$, $\Omega^{(1)}$ and
$\Delta^{(1)}$ is deferred until Section~\ref{evaluation}.  At present
it is sufficient to note that they possess Fourier expansions of the
form
\begin{equation}
  R^{(1)}(\lambda,\phi)={\rm Re}\sum_{m=0}^\infty
  R^{(1)(m)}(\lambda)\,{\rm e}^{{\rm i}m\phi},\qquad
  \Omega^{(1)}(\lambda,\phi)={\rm Re}\sum_{m=0}^\infty
  \Omega^{(1)(m)}(\lambda)\,{\rm e}^{{\rm i}m\phi},\qquad
  \Delta^{(1)}(\lambda,\phi)={\rm Re}\sum_{m=0}^\infty
  \Delta^{(1)(m)}(\lambda)\,{\rm e}^{{\rm i}m\phi},
\end{equation}
where $R^{(1)(m)}$ and $\Omega^{(1)(m)}$ are real owing to the
symmetry of the orbits, while $\Delta^{(1)(m)}$ is imaginary.

The assumption that the tidal potential scales as $O(\epsilon)$ is a
formal device that will turn out to provide a critical level of
non-linearity in the response.  Similarly, it is expedient to choose
viscosity parameters that scale formally as $O(\epsilon^{2/3})$, i.e.
\begin{equation}
  \alpha=\epsilon^{2/3}\tilde\alpha,\qquad
  \alpha_{\rm b}=\epsilon^{2/3}\tilde\alpha_{\rm b}.
\end{equation}
This will turn out to provide a critical level of dissipation in the
response.

\subsection{Expansion about the resonant orbit}

Let $\lambda=\lambda_*$ be the resonant orbit, to be identified
subsequently, and let a subscript $*$ generally denote a quantity
evaluated on the resonant orbit.  It is known from Lubow (1981) that
the characteristic radial extent of the resonance is $(H^2R)^{1/3}$,
as for a Lindblad resonance.  One therefore introduces a scaled
quasi-radial coordinate $\xi$ according to
\begin{equation}
  \lambda=\lambda_*+\epsilon^{2/3}\xi,
\end{equation}
so that $\xi=O(1)$ in the resonant region.  Any function of $\lambda$
may then be expanded in a Taylor series about the resonant orbit.  Thus
\begin{equation}
  R=\lambda_*+\epsilon^{2/3}\xi+\epsilon R_1(\phi)+O(\epsilon^{5/3}),
\end{equation}
\begin{equation}
  \Omega=\Omega_*+\epsilon^{2/3}\Omega_{2/3}\xi+\epsilon\Omega_1(\phi)+
  O(\epsilon^{4/3}),
\end{equation}
\begin{equation}
  \Delta=\epsilon\Delta_1(\phi)+O(\epsilon^{5/3}),
\end{equation}
where
\begin{equation}
  \Omega_*=\left({{GM_1}\over{\lambda_*^3}}\right)^{1/2}-\omega,
\end{equation}
\begin{equation}
  \Omega_{2/3}=-{{3}\over{2\lambda_*}}
  \left({{GM_1}\over{\lambda_*^3}}\right)^{1/2},
\end{equation}
while $R_1(\phi)=R^{(1)}(\lambda_*,\phi)$,
$\Omega_1(\phi)=\Omega^{(1)}(\lambda_*,\phi)$ and
$\Delta_1(\phi)=\Delta^{(1)}(\lambda_*,\phi)$.
Similarly,
\begin{equation}
  \Phi^{(2)}=\Phi^{(2)}_*+\epsilon^{2/3}\Phi^{(2)}_{2/3}\xi+
  \epsilon\Phi^{(2)}_1(\phi)+O(\epsilon^{4/3}),
\end{equation}
etc., where $\Phi^{(2)}_*=\tilde\Omega_*^2$ and
$\Phi^{(2)}_{2/3}=2\tilde\Omega_*\Omega_{2/3}$.

\subsection{Fluid variables}

A steady solution of the basic equations is sought.  After careful
analysis, the following expansion scheme for the fluid variables in
the resonant region has been found to be appropriate and
self-consistent.
\begin{eqnarray}
  u^\lambda&=&\epsilon^{5/3}u^\lambda_{5/3}(\xi,\phi,\zeta)+
  O(\epsilon^2),\\
  u^\phi-\Omega&=&\epsilon^{5/3}u^\phi_{5/3}(\xi,\phi,\zeta)+O(\epsilon^2),\\
  u^z&=&\epsilon^{4/3}u^z_{4/3}(\xi,\phi,\zeta)+
  \epsilon^{5/3}u^z_{5/3}(\xi,\phi,\zeta)+\epsilon^2u^z_2(\xi,\phi,\zeta)+
  O(\epsilon^{7/3}),\\
  \rho&=&\epsilon^s\left[\rho_*(\zeta)+
  \epsilon^{1/3}\rho_{1/3}(\xi,\phi,\zeta)+
  \epsilon^{2/3}\rho_{2/3}(\xi,\phi,\zeta)+\epsilon\rho_1(\xi,\phi,\zeta)+
  O(\epsilon^{4/3})\right],\\
  p&=&\epsilon^{s+2}\left[p_*(\zeta)+
  \epsilon^{1/3}p_{1/3}(\xi,\phi,\zeta)+
  \epsilon^{2/3}p_{2/3}(\xi,\phi,\zeta)+\epsilon p_1(\xi,\phi,\zeta)+
  O(\epsilon^{4/3})\right],\\
  \mu&=&\epsilon^{s+8/3}\left[\mu_*(\zeta)+
  \epsilon^{1/3}\mu_{1/3}(\xi,\phi,\zeta)+O(\epsilon^{2/3})\right],\\
  \mu_{\rm b}&=&\epsilon^{s+8/3}\left[\mu_{{\rm b}*}(\zeta)+
  O(\epsilon^{1/3})\right],\\
  T&=&\epsilon^2\left[T_*(\zeta)+\epsilon^{1/3}T_{1/3}(\xi,\phi,\zeta)+
  O(\epsilon^{2/3})\right],\\
  F^z&=&\epsilon^{s+11/3}\left[F^z_*(\zeta)+
  \epsilon^{1/3}F^z_{1/3}(\xi,\phi,\zeta)+
  O(\epsilon^{2/3})\right].
\end{eqnarray}
Here $s$ is a positive parameter, which drops out of the analysis,
although formally one requires $s=(10-6y)/(6+3x)$ in order to balance
powers of $\epsilon$ in the opacity law.  Other components of $F^a$
are smaller and will not be required.

According to this scheme, the relative amplitude of the perturbation
in the resonant region is $O(\epsilon^{1/3})$.  This is the magnitude
of the fractional perturbations of density and pressure, and of the
Mach number of the vertical velocity perturbation.  The relative
amplitude of the perturbation is greater than the relative magnitude
of the tidal potential, $O(\epsilon)$, because of the effect of the
resonance.  The horizontal velocity perturbations are smaller than the
vertical one by a factor $O(\epsilon^{1/3})$.

\subsection{Equations to be solved}

The above expansions for the fluid variables are now substituted into
the basic equations, which are expanded only as far as is required for
the analysis that follows.

\subsubsection{Equation of mass conservation}

$O(\epsilon^{s+1/3})$:
\begin{equation}
  \Omega_*\partial_\phi\rho_{1/3}+u^z_{4/3}\partial_\zeta\rho_*=
  -\rho_*\partial_\zeta u^z_{4/3}.
  \label{mass_1/3}
\end{equation}
$O(\epsilon^{s+2/3})$:
\begin{equation}
  \Omega_*\partial_\phi\rho_{2/3}+u^z_{4/3}\partial_\zeta\rho_{1/3}+
  u^z_{5/3}\partial_\zeta\rho_*=
  -\rho_*\partial_\zeta u^z_{5/3}-\rho_{1/3}\partial_\zeta u^z_{4/3}.
  \label{mass_2/3}
\end{equation}
$O(\epsilon^{s+1})$:
\begin{equation}
  \Omega_*\partial_\phi\rho_1+u^z_{4/3}\partial_\zeta\rho_{2/3}+
  (\Omega_{2/3}\xi\partial_\phi+u^z_{5/3}\partial_\zeta)\rho_{1/3}+
  u^z_2\partial_\zeta\rho_*=
  -\rho_*(\partial_\xi u^\lambda_{5/3}+\partial_\zeta u^z_2+\Delta_1)-
  \rho_{1/3}\partial_\zeta u^z_{5/3}-\rho_{2/3}\partial_\zeta u^z_{4/3}.
  \label{mass_1}
\end{equation}

\subsubsection{Energy equation}

$O(\epsilon^{s+7/3})$:
\begin{equation}
  \Omega_*\partial_\phi p_{1/3}+u^z_{4/3}\partial_\zeta p_*=
  -\gamma p_*\partial_\zeta u^z_{4/3}.
  \label{energy_7/3}
\end{equation}
$O(\epsilon^{s+8/3})$:
\begin{equation}
  \Omega_*\partial_\phi p_{2/3}+u^z_{4/3}\partial_\zeta p_{1/3}+
  u^z_{5/3}\partial_\zeta p_*=
  -\gamma p_*\partial_\zeta u^z_{5/3}-\gamma p_{1/3}\partial_\zeta u^z_{4/3}+
  (\gamma-1)\left(\mu_*\lambda_*^2\Omega_{2/3}^2-\partial_\zeta F^z_*\right).
  \label{energy_8/3}
\end{equation}
$O(\epsilon^{s+3})$:
\begin{eqnarray}
  \lefteqn{\Omega_*\partial_\phi p_1+u^z_{4/3}\partial_\zeta p_{2/3}+
  (\Omega_{2/3}\xi\partial_\phi+u^z_{5/3}\partial_\zeta)p_{1/3}+
  u^z_2\partial_\zeta p_*=
  -\gamma p_*(\partial_\xi u^\lambda_{5/3}+\partial_\zeta u^z_2+\Delta_1)-
  \gamma p_{1/3}\partial_\zeta u^z_{5/3}-
  \gamma p_{2/3}\partial_\zeta u^z_{4/3}}&\nonumber\\
  &&+(\gamma-1)\left(\mu_{1/3}\lambda_*^2\Omega_{2/3}^2-
  \partial_\zeta F^z_{1/3}\right).
  \label{energy_3}
\end{eqnarray}

\subsubsection{Equation of motion}

$\lambda$-component, $O(\epsilon^{s+5/3})$:
\begin{equation}
  \rho_*(\Omega_*\partial_\phi u^\lambda_{5/3}-
  2\lambda_*\tilde\Omega_*u^\phi_{5/3})=
  -\partial_\xi p_{1/3}.
  \label{ulambda_5/3}
\end{equation}
$\phi$-component, $O(\epsilon^{s+5/3})$:
\begin{equation}
  \rho_*\left(u^\lambda_{5/3}\Omega_{2/3}+
  \Omega_*\partial_\phi u^\phi_{5/3}+
  {{2\tilde\Omega_*}\over{\lambda_*}}u^\lambda_{5/3}\right)=0.
  \label{uphi_5/3}
\end{equation}
$z$-component, $O(\epsilon^{s+1})$:
\begin{equation}
  0=-\rho_*\Phi^{(2)}_*\zeta-\partial_\zeta p_*.
  \label{uz_1}
\end{equation}
$z$-component, $O(\epsilon^{s+4/3})$:
\begin{equation}
  \rho_*\Omega_*\partial_\phi u^z_{4/3}=-\rho_{1/3}\Phi^{(2)}_*\zeta-
  \partial_\zeta p_{1/3}.
  \label{uz_4/3}
\end{equation}
$z$-component, $O(\epsilon^{s+5/3})$:
\begin{equation}
  \rho_*\left(\Omega_*\partial_\phi u^z_{5/3}+
  u^z_{4/3}\partial_\zeta u^z_{4/3}\right)+
  \rho_{1/3}\Omega_*\partial_\phi u^z_{4/3}=-\rho_{2/3}\Phi^{(2)}_*\zeta-
  \rho_*\Phi^{(2)}_{2/3}\xi\zeta-\partial_\zeta p_{2/3}.
  \label{uz_5/3}
\end{equation}
$z$-component, $O(\epsilon^{s+2})$:
\begin{eqnarray}
  \lefteqn{\rho_*\left[\Omega_*\partial_\phi u^z_2+
  u^z_{4/3}\partial_\zeta u^z_{5/3}+
  (\Omega_{2/3}\xi\partial_\phi+u^z_{5/3}\partial_\zeta)u^z_{4/3}\right]+
  \rho_{1/3}\left(\Omega_*\partial_\phi u^z_{5/3}+
  u^z_{4/3}\partial_\zeta u^z_{4/3}\right)+
  \rho_{2/3}\Omega_*\partial_\phi u^z_{4/3}}&\nonumber\\
  &&=-\rho_1\Phi^{(2)}_*\zeta-
  \rho_{1/3}\Phi^{(2)}_{2/3}\xi\zeta-\rho_*\Phi^{(2)}_1\zeta-
  \rho_*\Phi^{(3)}_{{\rm t}*}\zeta-\partial_\zeta
  \left[p_1-(\mu_{{\rm b}*}+{\textstyle{{4}\over{3}}}\mu_*)
  \partial_\zeta u^z_{4/3}\right].
  \label{uz_2}
\end{eqnarray}

\subsubsection{Constitutive equations}

\begin{equation}
  F^z_*=-{{16\sigma T_*^{3-y}}\over{3C_\kappa\rho_*^{1+x}}}
  \partial_\zeta T_*,
\end{equation}
\begin{equation}
  p_*={{k\rho_*T_*}\over{\mu_{\rm m}m_{\rm H}}},
\end{equation}
\begin{equation}
  \mu_*={{\tilde\alpha p_*}\over{\tilde\Omega_*}},\qquad
  \mu_{{\rm b}*}={{\tilde\alpha_{\rm b}p_*}\over{\tilde\Omega_*}}.
\end{equation}
Also
\begin{equation}
  {{F^z_{1/3}}\over{F^z_*}}=(3-y){{T_{1/3}}\over{T_*}}-
  (1+x){{\rho_{1/3}}\over{\rho_*}}+
  {{\partial_\zeta T_{1/3}}\over{\partial_\zeta T_*}},
\end{equation}
\begin{equation}
  {{p_{1/3}}\over{p_*}}={{\rho_{1/3}}\over{\rho_*}}+{{T_{1/3}}\over{T_*}},
\end{equation}
\begin{equation}
  \mu_{1/3}={{\tilde\alpha p_{1/3}}\over{\tilde\Omega_*}}.
\end{equation}

\subsection{Solution}

\subsubsection{Unperturbed vertical structure}

From equation (\ref{uz_1}), one has the important relation
\begin{equation}
  \partial_\zeta p_*=-\rho_*\tilde\Omega_*^2\zeta,
\end{equation}
which expresses the vertical hydrostatic equilibrium of the
unperturbed disc.  The final two terms in equation (\ref{energy_8/3})
may be taken to balance, since these represent the unperturbed thermal
equilibrium of the disc, i.e.
\begin{equation}
  \partial_\zeta F^z_*={{9}\over{4}}\tilde\alpha p_*\tilde\Omega_*.
\end{equation}

\subsubsection{Form of the resonant mode}

Equations (\ref{mass_1/3}), (\ref{energy_7/3}) and (\ref{uz_4/3}) may
be combined to give
\begin{equation}
  \sL u^z_{4/3}=0,
  \label{luz_4/3}
\end{equation}
where
\begin{equation}
  \sL=\rho_*(\Omega_*^2\partial_\phi^2+\tilde\Omega_*^2)-
  \partial_\zeta(\gamma p_*\partial_\zeta)
\end{equation}
is a linear operator.  Normally $\sL$ would be a non-singular operator
and equation (\ref{luz_4/3}) would have no non-trivial solution.
However, the resonant orbit is, by definition, one on which $\sL$ is a
singular operator possessing a null eigenfunction of the form
\begin{equation}
  w={\rm e}^{{\rm i}m\phi}\,\zeta,
 \label{null}
\end{equation}
such that $\sL w=0$.  This requires
\begin{equation}
  m^2\Omega_*^2=(\gamma+1)\tilde\Omega_*^2,
  \label{resonance}
\end{equation}
which is the condition given by Lubow (1981) for a vertical resonance for
azimuthal wavenumber $m$ ($m>0$ always).  Accordingly, the general
solution of equation (\ref{luz_4/3}) is
\begin{equation}
  u^z_{4/3}={\rm Re}\left[-{\rm e}^{{\rm i}m\phi}\,
  {\rm i}m\Omega_*f\zeta\right]
\end{equation}
where $f(\xi)$ is a dimensionless function to be determined.  The
corresponding solutions of equations (\ref{mass_1/3}) and
(\ref{energy_7/3}) are
\begin{equation}
  \rho_{1/3}={\rm Re}\left[{\rm e}^{{\rm i}m\phi}\,
  f(\rho_*+\zeta\partial_\zeta\rho_*)\right],
\end{equation}
\begin{equation}
  p_{1/3}={\rm Re}\left[{\rm e}^{{\rm i}m\phi}\,
  f(\gamma p_*+\zeta\partial_\zeta p_*)\right].
\end{equation}
One also finds
\begin{equation}
  T_{1/3}={\rm Re}\left\{{\rm e}^{{\rm i}m\phi}\,
  f\left[(\gamma-1)T_*+\zeta\partial_\zeta T_*\right]\right\},
\end{equation}
\begin{equation}
  F^z_{1/3}={\rm Re}\left\{{\rm e}^{{\rm i}m\phi}\,
  f\left[((4-y)(\gamma-1)-x)F^z_*+\zeta\partial_\zeta F^z_*\right]\right\}.
\end{equation}
This motion represents a simple `breathing mode' of the disc.  In the
notation of Ogilvie (1998) it is the ${\rm p}_1^{\rm e}$ mode, i.e.
the first p mode of even symmetry about the mid-plane.

\subsubsection{Solvability condition for the linear operator}

At higher orders one naturally obtains equations of the form
\begin{equation}
  \sL X=F,
\end{equation}
where $X$ is unknown and $F$ is known.  Now $\sL$ is a singular
operator, and it is important to understand the conditions under which
such an equation is soluble.  Since $\sL$ is self-adjoint, the
solvability condition is easily obtained in the form
\begin{equation}
  \int\!\!\!\int w^*F\,{\rm d}\phi\,{\rm d}\zeta=0,
\end{equation}
where $w$ is the null eigenfunction (\ref{null}), and the integration is
over the full azimuthal and vertical extent of the disc.

\subsubsection{Horizontal velocities}

Equations (\ref{ulambda_5/3}) and (\ref{uphi_5/3}) may be combined to give
\begin{equation}
  \rho_*(\Omega_*^2\partial_\phi^2+\tilde\Omega_*^2)u^\lambda_{5/3}=
  -\Omega_*\partial_\phi\partial_\xi p_{1/3}.
\end{equation}
The solution is
\begin{equation}
  u^\lambda_{5/3}={\rm Re}\left[{\rm e}^{{\rm i}m\phi}\,
  {{{\rm i}m\Omega_*}\over{\gamma\tilde\Omega_*^2}}
  {{{\rm d}f}\over{{\rm d}\xi}}{{1}\over{\rho_*}}
  (\gamma p_*+\zeta\partial_\zeta p_*)\right],
\end{equation}
\begin{equation}
  u^\phi_{5/3}={\rm Re}\left[-{\rm e}^{{\rm i}m\phi}\,
  {{1}\over{2\gamma\lambda_*\tilde\Omega_*}}
  {{{\rm d}f}\over{{\rm d}\xi}}{{1}\over{\rho_*}}
  (\gamma p_*+\zeta\partial_\zeta p_*)\right],
\end{equation}
where equation (\ref{resonance}) has been used.

\subsubsection{Excitation of the second harmonic}

Equations (\ref{mass_2/3}), (\ref{energy_8/3}) and (\ref{uz_5/3}) may
be combined to give
\begin{equation}
  \sL u^z_{5/3}=F_{5/3},
  \label{luz_5/3}
\end{equation}
where 
\begin{equation}
  F_{5/3}=-\Omega_*\partial_\phi
  (\rho_*u^z_{4/3}\partial_\zeta u^z_{4/3}+
  \rho_{1/3}\Omega_*\partial_\phi u^z_{4/3})+
  \tilde\Omega_*^2\zeta
  (u^z_{4/3}\partial_\zeta\rho_{1/3}+
  \rho_{1/3}\partial_\zeta u^z_{4/3})+
  \partial_\zeta(u^z_{4/3}\partial_\zeta p_{1/3}+
  \gamma p_{1/3}\partial_\zeta u^z_{4/3}).
\end{equation}
This evaluates to
\begin{equation}
  F_{5/3}={\rm Re}\left[{\rm e}^{2{\rm i}m\phi}\,
  {{1}\over{2}}(\gamma+1)(\gamma+3){\rm i}m\Omega_*f^2
  \rho_*\tilde\Omega_*^2\zeta\right].
\end{equation}
The solvability condition is satisfied, and the solution is
\begin{equation}
  u^z_{5/3}={\rm Re}\left[-{\rm e}^{2{\rm i}m\phi}\,
  {{1}\over{6}}(\gamma+3){\rm i}m\Omega_*f^2\zeta\right].
\end{equation}
The corresponding solutions of equations (\ref{mass_2/3}) and
(\ref{energy_8/3}) are
\begin{equation}
  \rho_{2/3}={\rm Re}\left\{{\rm e}^{{2\rm i}m\phi}\,
  {{1}\over{4}}f^2\left[{{1}\over{3}}(\gamma+3)
  (\rho_*+\zeta\partial_\zeta\rho_*)+
  (\rho_*+\zeta\partial_\zeta\rho_*)+
  \zeta\partial_\zeta(\rho_*+\zeta\partial_\zeta\rho_*)\right]\right\}+
  \rho_{2/3}^{\rm ax}(\xi,\zeta),
\end{equation}
\begin{equation}
  p_{2/3}={\rm Re}\left\{{\rm e}^{{2\rm i}m\phi}\,
  {{1}\over{4}}f^2\left[{{1}\over{3}}(\gamma+3)
  (\gamma p_*+\zeta\partial_\zeta p_*)+
  \gamma(\gamma p_*+\zeta\partial_\zeta p_*)+
  \zeta\partial_\zeta(\gamma p_*+\zeta\partial_\zeta p_*)\right]\right\}+
  p_{2/3}^{\rm ax}(\xi,\zeta),
\end{equation}
where the axisymmetric parts satisfy
\begin{equation}
  0=-\rho_{2/3}^{\rm ax}\Phi^{(2)}_*\zeta-\rho_*\Phi^{(2)}_{2/3}\xi\zeta-
  \partial_\zeta p_{2/3}^{\rm ax}-
  {{1}\over{2}}(\gamma+1)\tilde\Omega_*^2|f|^2\partial_\zeta(\rho_*\zeta^2),
\end{equation}
and represent the Taylor expansion of the unperturbed vertical
structure about the resonant orbit, plus a non-linear term due to the wave.

\subsubsection{Excitation of the resonant mode}

Equations (\ref{mass_1}), (\ref{energy_3}) and (\ref{uz_2}) may
be combined to give
\begin{equation}
  \sL u^z_2=F_2,
  \label{luz_2}
\end{equation}
where
\begin{eqnarray}
  \lefteqn{F_2=-\Omega_*\partial_\phi
  \left\{\rho_*\left[u^z_{4/3}\partial_\zeta u^z_{5/3}+
  (\Omega_{2/3}\xi\partial_\phi+u^z_{5/3}\partial_\zeta)u^z_{4/3}\right]+
  \rho_{1/3}(\Omega_*\partial_\phi u^z_{5/3}+
  u^z_{4/3}\partial_\zeta u^z_{4/3})+
  \rho_{2/3}\Omega_*\partial_\phi u^z_{4/3}\right.}&\nonumber\\
  &&\left.\qquad\qquad+
  \rho_{1/3}\Phi^{(2)}_{2/3}\xi\zeta+
  \rho_*\Phi^{(2)}_1\zeta+\rho_*\Phi^{(3)}_{{\rm t}*}\zeta-
  \partial_\zeta\left[(\mu_{{\rm b}*}+{\textstyle{{4}\over{3}}}\mu_*)
  \partial_\zeta u^z_{4/3}\right]\right\}\nonumber\\
  &&+
  \tilde\Omega_*^2\zeta
  \left[u^z_{4/3}\partial_\zeta\rho_{2/3}+
  (\Omega_{2/3}\xi\partial_\phi+u^z_{5/3}\partial_\zeta)\rho_{1/3}+
  \rho_*(\partial_\xi u^\lambda_{5/3}+\Delta_1)+
  \rho_{1/3}\partial_\zeta u^z_{5/3}+
  \rho_{2/3}\partial_\zeta u^z_{4/3}\right]\nonumber\\
  &&+
  \partial_\zeta
  \left[u^z_{4/3}\partial_\zeta p_{2/3}+
  (\Omega_{2/3}\xi\partial_\phi+u^z_{5/3}\partial_\zeta)p_{1/3}+
  \gamma p_*(\partial_\xi u^\lambda_{5/3}+\Delta_1)+
  \gamma p_{1/3}\partial_\zeta u^z_{5/3}+
  \gamma p_{2/3}\partial_\zeta u^z_{4/3}\right.\nonumber\\
  &&\left.\qquad\qquad-
  (\gamma-1)\left(\mu_{1/3}\lambda_*^2\Omega_{2/3}^2-
  \partial_\zeta F^z_{1/3}\right)\right].
\end{eqnarray}
This equation need not be solved in detail.  It is sufficient to apply
the solvability condition,
\begin{equation}
  \int\!\!\!\int{\rm e}^{-{\rm i}m\phi}\,\zeta\,F_2
  \,{\rm d}\phi\,{\rm d}\zeta=0,
\end{equation}
which yields the required equation for $f(\xi)$.  After multiplication
by ${\rm i}/\pi$, this has the form
\begin{equation}
  C_1{{{\rm d}^2f}\over{{\rm d}\xi^2}}-
  C_2\xi f-{\rm i}C_3f-C_4|f|^2f+C_5=0,
  \label{cnla_primitive}
\end{equation}
where the five real, constant coefficients are given by
\begin{equation}
  C_1={{m\Omega_*}\over{\gamma\tilde\Omega_*^2}}
  \int{{1}\over{\rho_*}}(\gamma p_*+\zeta\partial_\zeta p_*)^2\,{\rm d}\zeta,
\end{equation}
\begin{equation}
  C_2=2(\gamma+1)m\tilde\Omega_*(\Omega_*-\tilde\Omega_*)\Omega_{2/3}
  \int\rho_*\zeta^2\,{\rm d}\zeta,
\end{equation}
\begin{equation}
  C_3=\left\{(\gamma+1)
  (\tilde\alpha_{\rm b}+{\textstyle{{4}\over{3}}}\tilde\alpha)+
  {{9}\over{4}}(\gamma-1)\left[(3-y)(\gamma-1)-x\right]\tilde\alpha\right\}
  \tilde\Omega_*^3\int\rho_*\zeta^2\,{\rm d}\zeta,
\end{equation}
\begin{equation}
  C_4={{1}\over{12}}\gamma(\gamma+1)(3-\gamma)m\Omega_*
  \tilde\Omega_*^2\int\rho_*\zeta^2\,{\rm d}\zeta,
\end{equation}
\begin{equation}
  C_5=m\Omega_*
  \left[\Phi^{(2)(m)}_1+\Phi^{(3)(m)}_{{\rm t}*}\right]
  \int\rho_*\zeta^2\,{\rm d}\zeta+
  (\gamma-1)\left(-{\rm i}\Delta_1^{(m)}\right)
  \tilde\Omega_*^2\int\rho_*\zeta^2\,{\rm d}\zeta.
\end{equation}

Equation (\ref{cnla_primitive}) may be reduced to a {\it normal form},
\begin{equation}
  -{{{\rm d}^2y}\over{{\rm d}x^2}}+xy+|y|^2y+{\rm i}by=a,
  \label{cnla}
\end{equation}
by means of the transformations
\begin{equation}
  x=C_1^{-1/3}C_2^{1/3}\xi,
\end{equation}
\begin{equation}
  y(x)=C_1^{-1/6}C_2^{-1/3}C_4^{1/2}f(\xi).
\end{equation}
The two essential parameters are
\begin{equation}
  a=C_1^{-1/2}C_2^{-1}C_4^{1/2}C_5,
\end{equation}
\begin{equation}
  b=C_1^{-1/3}C_2^{-2/3}C_3,
\end{equation}
and are both real.

\subsection{Relation to physical units}

When the asymptotic scalings are removed, the dimensionless parameters
can be expressed directly in terms of the physical variables evaluated
at the location of the resonance.  The ordering parameter $\epsilon$
then cancels out.  However, the integral involved in $C_1$ introduces
the dimensionless parameter $\hat\epsilon$ defined by
\begin{equation}
  \int{{1}\over{\rho}}
  \left(\gamma p+z{{\partial p}\over{\partial z}}\right)^2\,{\rm d}z=
  \hat\epsilon^2\lambda_*^2\tilde\Omega_*^4\int\rho z^2\,{\rm d}z.
\end{equation}
This is a more specific measure of $H/R$, and depends on the vertical
structure of the disc.  For a vertically isothermal disc,
$\hat\epsilon^2=(\gamma^2-2\gamma+3)(H/R)^2$, where $H$ is the
isothermal scale height.  For a vertically polytropic disc with
polytropic index $3/2$,
$\hat\epsilon^2=(1/40)(7\gamma^2-10\gamma+15)(H/R)^2$, where $H$ is
the true semi-thickness.  For a vertically polytropic disc with
polytropic index $3$,
$\hat\epsilon^2=(1/44)(5\gamma^2-8\gamma+12)(H/R)^2$.

A quantity deriving from $C_2$ is the rate of detuning of the
vertical resonance,
\begin{equation}
  \sD={{\rm d}\over{{\rm d}\ln\lambda}}
  \left[(\gamma+1)\tilde\Omega^2-m^2\Omega^2\right].
\end{equation}
Another quantity having the dimensions of frequency-squared is the
forcing combination appearing in $C_5$,
\begin{equation}
  \Psi=\Phi_1^{(2)(m)}+\Phi_{{\rm t}*}^{(3)(m)}+
  (\gamma-1)\left({{\Delta_1^{(m)}}\over{{\rm i}m\Omega_*}}\right)
  \tilde\Omega_*^2.
\end{equation}

One then finds
\begin{equation}
  a=\left[{{\gamma^2(\gamma+1)(3-\gamma)}\over{12}}\right]^{1/2}
  {{\Psi}\over{\hat\epsilon\sD}}.
\end{equation}
This shows why the tidal potential was chosen to scale formally as
$O(\epsilon)$, in order to have $a=O(1)$.  Similarly,
\begin{equation}
  b=\left\{(\gamma+1)
  (\alpha_{\rm b}+{\textstyle{{4}\over{3}}}\alpha)+
  {{9}\over{4}}(\gamma-1)\left[(3-y)(\gamma-1)-x\right]\alpha\right\}
  \left[{{\gamma\tilde\Omega_*^4}\over
  {(\gamma+1)^{3/2}\hat\epsilon^2\sD^2}}\right]^{1/3}.
\end{equation}
The scaling between $x$ and the physical length $\lambda$ is given by
\begin{equation}
  {{{\rm d}\lambda}\over{{\rm d}x}}=
  \left({{\hat\epsilon^2\tilde\Omega_*^2}\over{\gamma\sD}}\right)^{1/3}
  \lambda_*.
\end{equation}

\subsection{Evaluation of the forcing terms}

\label{evaluation}

There are three contributions to the forcing coefficient $\Psi$.  The
direct vertical forcing involves two terms: $\Phi_1^{(2)(m)}$, which
is the perturbation of the primary potential due to the tidal
distortion of the orbit, and $\Phi_{{\rm t}*}^{(3)(m)}$, which is the tidal
potential itself.  The term proportional to $\Delta_1^{(m)}$ is an indirect
forcing of the p mode resulting from the divergence of the tidally
distorted orbital motion.

In the following it is assumed that the resonance is an inner vertical
resonance with $m\ge2$.  The Fourier components of the tidal potential
can be expressed in terms of the Laplace coefficients
\begin{equation}
  b_\gamma^{(m)}(\beta)={{2}\over{\pi}}\int_0^\pi
  (1+\beta^2-2\beta\cos\phi)^{-\gamma}\cos(m\phi)\,{\rm d}\phi,
\end{equation}
where $\beta=\lambda/d$.  Thus
\begin{equation}
  \Phi_{\rm t}^{(1)(m)}=-{{GM_2}\over{d}}\,b_{1/2}^{(m)}(\beta),\qquad
  \Phi_{\rm t}^{(3)(m)}={{GM_2}\over{d^3}}\,b_{3/2}^{(m)}(\beta).
\end{equation}
The location of the resonance is
\begin{equation}
  {{\lambda_*}\over{d}}=\beta_*=(1+q)^{-1/3}
  \left[1-{{1}\over{m}}(\gamma+1)^{1/2}\right]^{2/3}.
\end{equation}

The orbital equations (\ref{orbital_r}) and (\ref{orbital_phi}) yield
the following linear equations for the perturbed orbital quantities:
\begin{equation}
  \left(m^2\Omega^{(0)2}+3\tilde\Omega^2\right)\lambda R^{(1)(m)}+
  2\tilde\Omega\lambda^2\Omega^{(1)(m)}=
  \lambda{{\rm d}\over{{\rm d}\lambda}}\Phi_{\rm t}^{(1)(m)},
\end{equation}
\begin{equation}
  \lambda\Omega^{(0)}\left(\lambda\Omega^{(1)(m)}+
  2\tilde\Omega R^{(1)(m)}\right)=-\Phi_{\rm t}^{(1)(m)}.
\end{equation}
Thus
\begin{equation}
  \left(m^2\Omega^{(0)2}-\tilde\Omega^2\right)\lambda R^{(1)(m)}=
  \lambda{{\rm d}\over{{\rm d}\lambda}}\Phi_{\rm t}^{(1)(m)}+
  {{2\tilde\Omega}\over{\Omega^{(0)}}}\Phi_{\rm t}^{(1)(m)}
\end{equation}
and
\begin{eqnarray}
  \Delta^{(1)(m)}&=&{\rm i}m\left[\Omega^{(1)(m)}+\Omega^{(0)}
  \left({{R^{(1)(m)}}\over{\lambda}}+
  {{\rm d}\over{{\rm d}\lambda}}R^{(1)(m)}\right)\right]\\
  &=&{\rm i}m\left[-{{1}\over{\lambda^2\Omega^{(0)}}}\Phi_{\rm t}^{(1)(m)}-
  {{2\tilde\Omega}\over{\lambda}}R^{(1)(m)}+\Omega^{(0)}
  \left({{R^{(1)(m)}}\over{\lambda}}+
  {{\rm d}\over{{\rm d}\lambda}}R^{(1)(m)}\right)\right].
\end{eqnarray}

Evaluating these quantities at the resonance, one finds
\begin{equation}
  \Phi_1^{(2)(m)}=-3{{GM_1}\over{\lambda_*^4}}R_1^{(m)},
\end{equation}
\begin{equation}
  \Delta_1^{(m)}={\rm i}m\left\{{{3}\over{\gamma\lambda_*}}
  \left[(\gamma+1)\tilde\Omega_*-\Omega_*\right]R_1^{(m)}-
  {{\Omega_*}\over{\gamma\tilde\Omega_*^2}}
  \left(\Phi_{{\rm t}*}^{(3)(m)}+
  {{3\tilde\Omega_*}\over{\lambda_*^2\Omega_*}}
  \Phi_{{\rm t}*}^{(1)(m)}\right)\right\},
\end{equation}
and therefore
\begin{eqnarray}
  \lefteqn{\Psi={{GM_1}\over{d^3}}{{q}\over{\gamma\beta_*^2}}
  \left\{\beta_*^2b_{3/2}^{(m)}(\beta_*)+
  3m(\gamma+1)^{-1/2}(\gamma-1)b_{1/2}^{(m)}(\beta_*)\right.}&\nonumber\\
  &&\qquad\qquad\qquad\qquad\qquad\qquad\qquad\left.
  +{{3}\over{\gamma}}\left[2\gamma-1-m(\gamma+1)^{1/2}(\gamma-1)\right]
  \left[\beta_*b_{1/2}^{(m)\prime}(\beta_*)+
  2m(\gamma+1)^{-1/2}b_{1/2}^{(m)}(\beta_*)\right]\right\}.
\end{eqnarray}
Here the condition for an inner vertical resonance,
\begin{equation}
  m\Omega_*=(\gamma+1)^{1/2}\tilde\Omega_*
\end{equation}
has been used, as well as the fact that the tidal potential satisfies
Laplace's equation,
\begin{equation}
  {{1}\over{\lambda}}{{\rm d}\over{{\rm d}\lambda}}
  \left(\lambda{{\rm d}\over{{\rm d}\lambda}}\Phi_{\rm t}^{(1)(m)}\right)
  -{{m^2}\over{\lambda^2}}\Phi_{\rm t}^{(1)(m)}+\Phi_{\rm t}^{(3)(m)}=0.
\end{equation}
One also finds
\begin{equation}
  \sD={{GM_1}\over{d^3}}\left[3m(\gamma+1)^{1/2}(1+q)^{1/2}\beta_*^{-3/2}
  \right].
\end{equation}

\subsection{Illustrative parameters}

\label{illustrative}

The following parameter values might be appropriate for a system such
as IP~Peg: $\gamma=5/3$, $m=2$ and $q=0.5$.  For the purposes of
illustration, suppose that $\alpha_{\rm b}=0$, $x=1$ and $y=-7/2$.
Then one finds
\begin{equation}
  \lambda_*\approx0.282\,d,\qquad
  a\approx0.0110\,\hat\epsilon^{-1},\qquad
  b\approx4.20\,\alpha\hat\epsilon^{-2/3},\qquad
  {{{\rm d}\lambda}\over{{\rm d}x}}\approx0.196\,\hat\epsilon^{2/3}d.
\end{equation}
For these parameters, radiative damping provides a somewhat greater
contribution to $b$ than does viscous damping.  A more non-linear and
less dissipative example occurs when $\gamma=6/5$, $m=2$ and $q=1$:
\begin{equation}
  \lambda_*\approx0.322\,d,\qquad
  a\approx0.0725\,\hat\epsilon^{-1},\qquad
  b\approx1.26\,\alpha\hat\epsilon^{-2/3},\qquad
  {{{\rm d}\lambda}\over{{\rm d}x}}\approx0.230\,\hat\epsilon^{2/3}d.
\end{equation}

\section{Analysis of the complex non-linear Airy equation}

\subsection{Physical interpretation}

The non-linear ordinary differential equation (\ref{cnla}) describes
the response of the disc in the neighbourhood of the vertical
resonance.  All the fluid variables can be determined from the
solution of this equation.  A detailed study of the solutions is
therefore called for.

Equation (\ref{cnla}) does not appear to have been studied before.  It
may be conveniently described as the {\it complex non-linear Airy
  equation}.  The five terms may be interpreted as follows.  The
linear terms $-y''+xy$ represent the ability of the
disc to support freely propagating waves in the region $x<0$.  In the
present case these waves correspond to the compressive ${\rm p}_1^{\rm
  e}$ mode.  The point $x=0$ is the location of the resonance, which
is also the turning point of free waves.  The non-linear term $|y|^2y$
derives from mode couplings.  The linear term ${\rm i}by$ represents
the dissipation of vertical motions by shear and bulk viscosity and by
radiative damping.  The term $a$ is the tidal forcing.

It is important to realize that the cubic form of the non-linear term
is {\it not\/} a low-order truncation of a more complicated
expression, but is the {\it exact\/} form of the non-linearity within
the consistent ordering scheme described above.  The cubic term arises
as follows.  For simplicity, consider only the vertical velocity
$u^z$, which is proportional to $z$.  In the equation for $u^z$ there
is a non-linear term $u^z\partial_zu^z$, also proportional to $z$,
that couples together different azimuthal modes.  With the scalings
adopted in this paper, the vertical tidal acceleration
$-\partial_z\Phi_{\rm t}$ is $O(\epsilon^2)$.  All azimuthal modes
respond to this forcing with $u^z=O(\epsilon^2)$, except the resonant
mode $m$, which responds with $u^z=O(\epsilon^{4/3})$.  The quadratic
self-interaction of mode $m$ forces modes $0$ and $2m$ at
$O(\epsilon^{5/3})$, i.e. at lower order than the tidal forcing, and
they respond with $u^z=O(\epsilon^{5/3})$.  The interaction of these
modes with mode $m$ forces modes $m$ and $3m$ at $O(\epsilon^2)$, i.e.
at the same order as the tidal forcing.  This modifies the
leading-order resonant mode, leading to a cubic non-linearity
(Fig.~1).

\begin{figure*}
  \centerline{\epsfbox{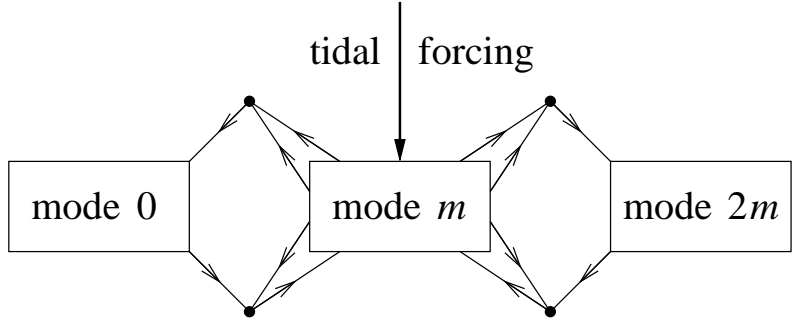}}
  \vskip0.5cm
  \caption{Mode couplings producing the cubic non-linearity.  Quadratic
    non-linear terms in the basic equations, such as
    $u^z\partial_zu^z$, couple together the different azimuthal mode
    numbers.  The combination of any two modes $m_1$ and $m_2$ (shown
    by arrows converging on a dot) forces modes $|m_1\pm m_2|$ (each
    shown by an arrow leaving a dot).  Although all modes are forced
    at similar levels by the tidal potential, the resonant mode $m$
    responds with anomalously large amplitude.  Its non-linear
    self-interaction excites modes $0$ and $2m$ at larger amplitudes
    than they would acquire by direct forcing, and their own
    interactions with mode $m$ feed back on mode $m$.  The net result
    is that mode $m$ affects itself through a cubic non-linearity.}
\end{figure*}

\subsection{Special limits}

When $a\ll1$, the response is $y=O(a)$ and the cubic term may be
neglected.  One then has
\begin{equation}
  -{{{\rm d}^2y}\over{{\rm d}x^2}}+xy+{\rm i}by=a.
  \label{cla}
\end{equation}
In the absence of dissipation, $b=0$ and one obtains an inhomogeneous
Airy equation,
\begin{equation}
  -{{{\rm d}^2y}\over{{\rm d}x^2}}+xy=a,
  \label{la}
\end{equation}
which is well known in studies of resonant wave excitation in discs.
The general solution is
\begin{equation}
  y=\pi a{\rm Gi}(x)+c_1{\rm Ai}(x)+c_2{\rm Bi}(x),
\end{equation}
where $c_1$ and $c_2$ are constants, ${\rm Ai}$ and ${\rm Bi}$ are the
Airy functions, and ${\rm Gi}$ is the inhomogeneous Airy function
defined by, e.g., Abramowitz \& Stegun (1965).  As $x\to+\infty$, both
${\rm Gi}(x)$ and ${\rm Ai}(x)$ decay monotonically, while ${\rm
  Bi}(x)$ diverges exponentially.  Therefore $c_2=0$ for an acceptable
solution.  For $x<0$, both ${\rm Gi}(x)$ and ${\rm Ai}(x)$ are
wavelike.  The desired solution is
\begin{equation}
  y=\pi a\left[{\rm Gi}(x)-{\rm i}{\rm Ai}(x)\right],
\end{equation}
which has the asymptotic form
\begin{equation}
  y\sim\pi^{1/2}a(-x)^{-1/4}\exp
  \left[-{{2}\over{3}}{\rm i}(-x)^{3/2}-{{{\rm i}\pi}\over{4}}\right]
\end{equation}
as $x\to-\infty$, and represents a purely outgoing wave emitted from
the resonance and travelling to the left.  For any other choice of
$c_1$, the solution would contain an incoming wave component that is
unphysical unless the outgoing wave is able to reach an edge of the
disc, reflect and return to the site of launching with a
non-negligible amplitude.

When dissipation is present so that $b>0$, the emitted wave is damped.
Formally, the solution may be obtained simply by replacing $x$ by
$x+{\rm i}b$.  This effectively moves the resonance off the real axis.

\subsection{Angular momentum flux and tidal torque}

One can multiply equation (\ref{cnla}) by $y^*$ and take the imaginary
part to deduce the conservation law
\begin{equation}
  {{\rm d}\over{{\rm d}x}}{\rm Im}
  \left(-y^*{{{\rm d}y}\over{{\rm d}x}}\right)=a\,{\rm Im}(-y)-b|y|^2.
\end{equation}
In the absence of dissipation and forcing, there is a strictly
conserved quantity which is proportional to the angular momentum flux
of the wave.  The term $a\,{\rm Im}(-y)$ is proportional to the tidal
torque density, and the term $-b|y|^2$ represents dissipative losses.
The total tidal torque exerted on the disc is therefore proportional
to
\begin{equation}
  a\int_{-\infty}^\infty{\rm Im}(-y)\,{\rm d}x.
\end{equation}
In the linear, inviscid limit ($a\ll1$, $b=0$) this quantity is equal
to
\begin{equation}
  \pi a^2\int_{-\infty}^\infty{\rm Ai}(x)\,{\rm d}x=\pi a^2.
\end{equation}
Define the {\it tidal torque parameter\/}
\begin{equation}
  f_T={{1}\over{\pi a}}\int_{-\infty}^\infty{\rm Im}(-y)\,{\rm d}x.
\end{equation}
In linear theory $f_T=1$.  When $b=0$ the torque goes entirely into
launching the wave.  When dissipation is included, it is still true
that $f_T=1$, because the contour integral is unchanged when the
function is shifted parallel to the imaginary axis.  The torque then
goes partly into launching the wave and partly directly into changing
the angular momentum of the disc.  As the wave attenuates, its angular
momentum is also transferred to the disc.  In non-linear theory
$f_T\ne1$ in general.

It can be shown that the tidal torque, in physical terms, is equal to
\begin{equation}
  T=-f_T{{\pi\lambda^2m\Psi^2}\over{\sD}}\int\rho z^2\,{\rm d}z,
  \label{torque}
\end{equation}
where all quantities are to be evaluated at the resonance
$\lambda=\lambda_*$.  This is closely related to the standard formula
for the torque exerted at a Lindblad resonance (Goldreich \& Tremaine
1979) except that (i) vertical rather than horizontal forcing is
involved, (ii) the second vertical moment of the density, rather than
the surface density, appears, and (iii) the denominator involves the
rate of detuning of the p-mode resonance condition rather than the
epicyclic resonance.  The factor $f_T$ also accounts for a non-linear
correction to the torque.  In the case $f_T=1$ and for an isothermal
disc with $\gamma=1$, this formula agrees with equation (40) of Lubow
(1981).

\subsection{Numerical method of solution}

In the non-linear case, equation (\ref{cnla}) must be solved
numerically.  In principle, it might be solved as a two-point
boundary-value problem with suitable conditions at large positive and
large negative $x$.  However, no satisfactory method was found to
solve the non-linear equation by this method owing to the existence of
unwanted growing solutions.

Instead, equation (\ref{cnla}) has been solved by converting it into a
time-dependent partial differential equation,
\begin{equation}
  {\rm i}{{\partial y}\over{\partial t}}-
  {{\partial^2y}\over{\partial x^2}}+xy+|y|^2y+{\rm i}by=a.
  \label{pde}
\end{equation}
The steady solutions of this equation are identical to the solutions
of equation (\ref{cnla}), and can be found numerically by solving the
time-dependent equation as an initial-value problem starting from a
suitable initial condition such as $y=0$.  In fact, this is not a
purely formal device.  It is possible to extend the analysis of this
paper to permit the solution to vary slowly in time in the binary
frame.  This can be done by allowing the perturbations to depend on a
slow time coordinate $\epsilon^{2/3}t$ in addition to $\xi$, $\phi$
and $\zeta$.  The resulting amplitude equation is precisely equation
(\ref{pde}), with a suitably scaled dimensionless time variable.
Therefore time-dependent solutions of equation (\ref{pde}) that
approach a steady state genuinely describe the development of a steady
wave pattern in an initially undisturbed disc, and also suggest the
stability of the steady solution.

Equation (\ref{pde}) is a dispersive, non-linear wave equation related
to the non-linear Schr\"odinger equation and the Ginzburg-Landau
equation.  The term $xy$ breaks the translational invariance,
providing a turning point for waves at $x=0$.  The term ${\rm i}by$
with $b>0$ ensures that all solutions decay in the absence of the
forcing term $a$.

Equation (\ref{pde}) has been solved by discretizing it in space on a
regular grid of 5000 points on the interval $-100<x<100$.  A
second-order, centred finite difference approximation of the second
derivative was used.  The desired solution is evanescent for large
positive $x$ and has the form of an outgoing wave at large negative
$x$.  In order to select this solution and to prevent wave reflection
as far as possible, the computational domain was extended to
$-110<x<110$ with an additional critical damping effect applied in the
added intervals.  The boundary condition $y=0$ was then applied at
$x=\pm110$.  The resulting ordinary differential equations in the
temporal domain were solved using a fifth-order Runge-Kutta method
with adaptive step-size.  Starting from the initial condition $y=0$,
the solution was advanced until a steady solution was obtained.  The
torque parameter was then evaluated by integrating the solution over
$-100<x<100$.

\begin{figure*}
  \centerline{\epsfbox{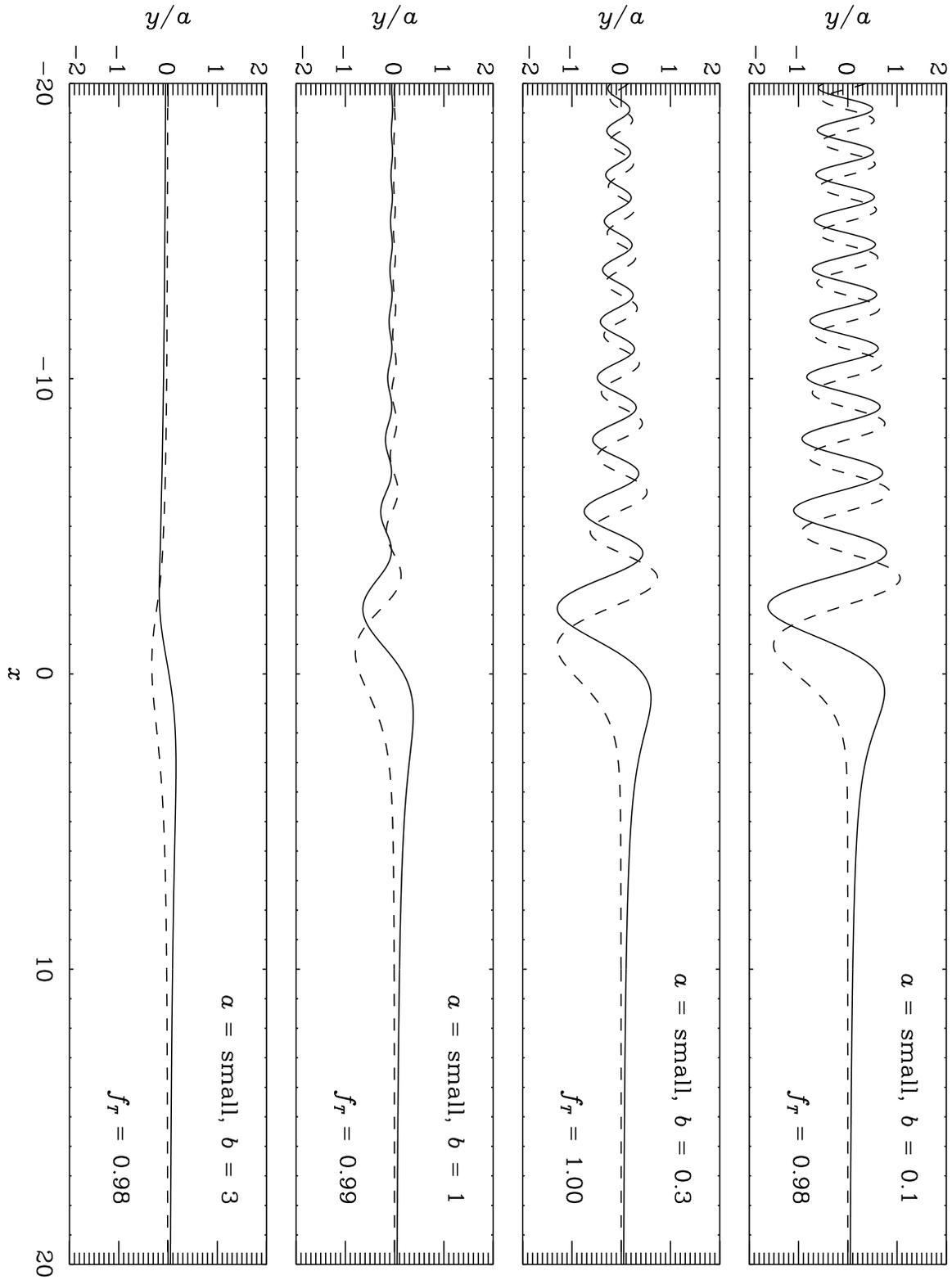}}
  \vskip-1cm
  \caption{Effect of increasing the dissipation parameter $b$ on solutions
    of the complex non-linear Airy equation.}
\end{figure*}

\begin{figure*}
  \centerline{\epsfbox{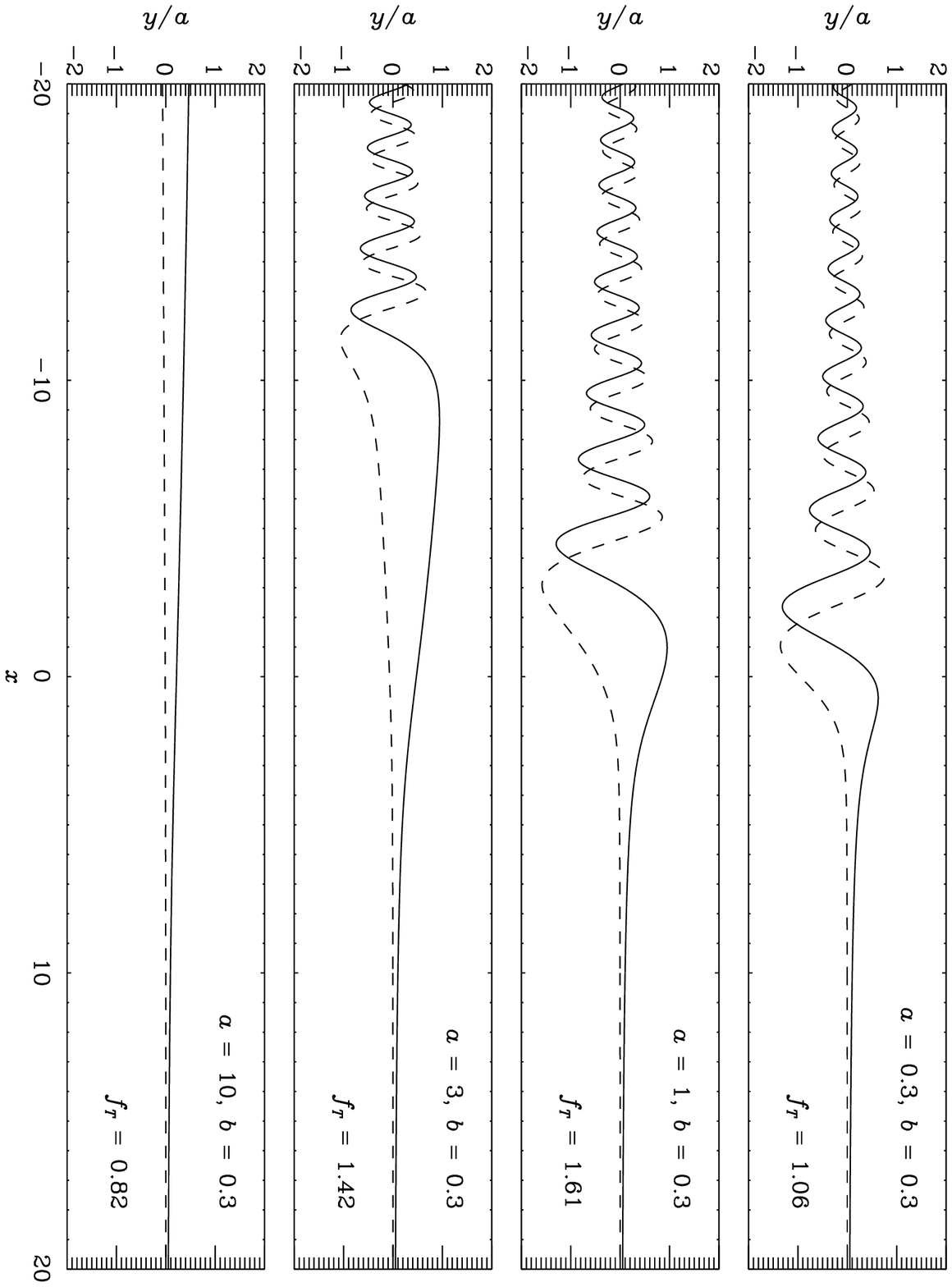}}
  \vskip-1cm
  \caption{Effect of increasing the amplitude parameter $a$ on solutions
    of the complex non-linear Airy equation.}
\end{figure*}

Some illustrative solutions of the complex non-linear Airy equation
are shown in Figs~2 and~3.  For each solution, the estimated value of
the torque parameter $f_T$ is quoted.  It should be borne in mind
that, for those solutions in which a wave of significant amplitude
reaches the edge of the computational domain at $x=-100$, the
integrated torque continues to fluctuate spatially at the level of a
few per cent, and so the value of $f_T$ depends to this extent on
exactly where the boundary is drawn.

In Fig.~2 the effect of increasing the dissipation parameter $b$ on
solutions in the linear regime $a\ll1$ is investigated.  For $b\la1$
the most noticeable effect is the attenuation of the emitted wave.
For $b\ga1$ the emitted wave is almost completely absent and the
resonant response is a broad, smooth distortion.  The estimated torque
parameters are consistent with the analytical expectation that $f_T=1$
when $a\ll1$, for any value of $b$.

In Fig.~3 the effect of increasing the amplitude parameter $a$ is
investigated.  The most obvious effect is a broadening of the
resonance, in which the first wavelength of the emitted wave becomes
distorted into a much wider feature.  The amplitude of the emitted
wave increases initially more rapidly than $a$, and $f_T$ achieves
values in excess of unity.  For larger $a$, however, $f_T$ declines to
values less than unity, perhaps indicating a saturation of the
resonance.

When either $a$ or $b$ is large, the resonance is substantially
broadened and the scaling assumptions used to derive the complex
non-linear Airy equation become inapplicable.  Under these conditions,
the vertical resonance can be treated, as in Paper~I, as a part of the
smooth tidal distortion of the disc.

\section{Discussion}

In this paper I have developed a general theory of vertical
resonances, first analysed by Lubow (1981), which are an important
aspect of the interaction between an accretion disc and a massive
companion with a coplanar orbit.  When Lubow's analysis is generalized
to allow for non-linearity of the response, and dissipation by
radiative damping and turbulent viscosity, the problem reduces to a
universal, non-linear ordinary differential equation with two real
parameters.

Numerical solutions of a time-dependent version of the complex
non-linear Airy equation describe the process by which a steady and
stable pattern of non-axisymmetric distortion is established in an
initially undisturbed disc.  For small values of the dissipation
parameter $b$, a p-mode wave is launched at the resonance and
propagates radially through the disc, carrying angular momentum.  For
larger values of $b$ the wave is attenuated or not launched at all,
and the resonant response is broadened into a smooth feature.  The
effect of increasing the amplitude parameter $a$ is to broaden the
resonance and ultimately to saturate it.  The total tidal torque
exerted at the resonance is independent of $b$ in the linear regime
$a\ll1$, but can be either greater or less than the linear prediction
when non-linearity is important.

If the magnitude of the resonant tidal torque exceeds that of the
local viscous torque, the disc may be truncated at the resonance
(Lubow 1981).  The viscous torque in the absence of tidal distortions
is
\begin{equation}
  \int\!\!\!\int\lambda^3\mu{{{\rm d}\Omega}\over{{\rm d}\lambda}}
  \,{\rm d}\phi\,{\rm d}z=
  -3\pi\alpha\lambda^2\tilde\Omega^2\int\rho z^2\,{\rm d}z,
\end{equation}
and a comparison with equation (\ref{torque}) shows that truncation
occurs if $\alpha<\alpha_{\rm trunc}$, where
\begin{equation}
  \alpha_{\rm trunc}={{f_Tm\Psi^2}\over{3\sD\tilde\Omega^2}}.
\end{equation}
However, if $\alpha$ is so small that the launched wave does not damp
significantly in the vicinity of the resonance, the tidal torque is
not transmitted to the disc locally and the process of truncation
cannot occur in a straightforward manner.  In the derivation of the
complex non-linear Airy equation it was assumed that
$\alpha=O(\epsilon^{2/3})$ while $\Psi=O(\epsilon)$.  Therefore the
process of tidal truncation is not described by that equation, but
occurs in a different parameter regime.

Some illustrative values of $\alpha_{\rm trunc}$ for the $m=2$ inner
vertical resonance are given in Table~1, where it is assumed that
$f_T=1$.  It appears that truncation by this resonance is unlikely in
cataclysmic variable discs, even in quiescence, and especially if
$\gamma\approx5/3$.  Although Lubow (1981) argued that truncation is
likely, his estimates were based on the case $\gamma=1$, for which the
resonant radius is maximal.  This results in a much larger torque than
is obtained when $\gamma=5/3$, because of the proximity to the
companion.  However, for the same reason, the resonance is less likely
to lie inside the disc in the case $\gamma=1$.

\begin{table*}
  \centering
  \caption{Critical value of the viscosity parameter below which the disc
    may be truncated at the $m=2$ inner vertical resonance.}
  \begin{tabular}{@{}lrr@{}}
  $q$&$\gamma$&$\alpha_{\rm trunc}$\\
  \hline
  1&1&0.07489\\
  &6/5&0.01694\\
  &5/3&0.00037\\
  0.5&1&0.03533\\
  &6/5&0.00796\\
  &5/3&0.00018\\
  0.2&1&0.00935\\
  &6/5&0.00210\\
  &5/3&0.00005
\end{tabular}
\end{table*}

Much more probable is that a non-destructive, two-armed structure is
formed, as described by the complex non-linear Airy equation.  In
Paper~I I have argued that this effect, in the limit in which the
resonant response is broadened and damped by non-linearity and
dissipation, could explain the non-axisymmetric structures seen in
Doppler tomograms of dwarf novae in outburst.  Indeed, estimates of
the parameters for a system such as IP~Peg
(Section~\ref{illustrative}) suggest that the resonance is strongly
damped and mildly non-linear.

It is likely that the techniques used in this paper will be useful in
other circumstances where spatial resonances occur in wave-bearing
media subject to periodic forcing.  The equation derived has a concise
form and its terms have a clear physical interpretation, suggesting
that it may be generic to some extent.  In any case it provides a
simple mathematical model in which the effects of non-linearity and
dissipation on the resonant launching of waves can be studied.

\section*{Acknowledgments}

I thank Jon Dawes, Steve Lubow and Jim Pringle for helpful
discussions.  I acknowledge the support of the Royal Society through a
University Research Fellowship.

\label{lastpage}

\end{document}